\documentclass{article}
\usepackage{hiph-art}
\newcommand{\be}{\begin{equation}}
\newcommand{\ee}{\end{equation}}
\newcommand{\ba}{\begin{eqnarray}}
\newcommand{\ea}{\end{eqnarray}}
\newcommand{\NL}{\nonumber \\}

\volnumber{19} \issuenumber{1} \edyear{2004}                             
\frompage{000} \topage{000}                                              
\recrevdate{6 Aprilis 2004}                                              
\def\rightmark{Quark coalescence into opposite parity ...}
\def\leftmark{J. Zim\'anyi and P. L\'evai}

\newcommand{\insertplotg}[1]{\begin{center}\leavevmode\epsfysize=10.5cm \epsfbox{#1}\end{center}}

\newcommand{\insertplotr}[1]{\begin{center}\leavevmode\epsfysize=5.5cm \epsfbox{#1}\end{center}}
\newcommand{\insertploth}[1]{\begin{center}\leavevmode\epsfysize=3.1cm \epsfbox{#1}\end{center}}
 
\title{Quark coalescence into opposite parity baryon states }
\authors{ 
{J\'ozsef Zim\'anyi$^1$ and P\'eter L\'evai$^{1}$ %
\index{Zim\'anyi, J} 
\index{L\'evai, P.} 
}\\[2.812mm]
{\normalsize
\hspace*{-8pt}$^1$ KFKI Research Institute for Nuclear and Particle Physics\\ 
49 PO. Box, Budapest, H-1525, Hungary 
}}
 
\abstract{
The production rate of
negative parity baryons was found to be much weaker than that of positive
states in RHIC experiments. In the present paper we show that this 
suppression is a simple
consequence of the coalescence dynamics of hadronization. }

 
\begin{document}
 
\maketitle
\setcounter{page}{1}

\section{Preludium}\label{prelud}

The speculations on the nature of quark gluon plasma (QGP) has a long 
history. It was assumed, that due to the large momenta of 
quarks and gluons in the plasma phase, the interaction between them
become negligible small as a consequence of the running coupling
constant.  
This  idea of non interacting  massless quarks and gluons
 in a big bag was used by many authors, e.g. the authors 
listed in Ref.~\cite{list}.

\begin{figure}[tp]
\setlength{\epsfxsize=0.95\textwidth}
\setlength{\epsfysize=0.47\textheight}
\centerline{\epsffile{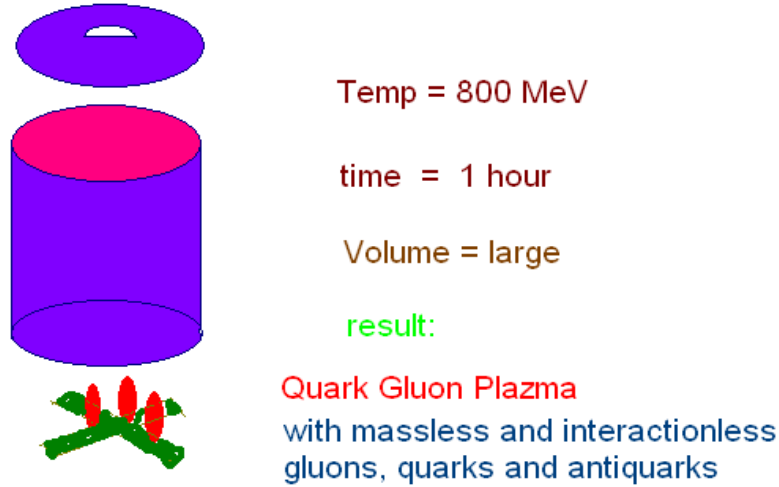}}
\caption[]{
Visualization of the cooking of  quark-gluon soup.
  }
\label{fig1}
\end{figure}

However, it became clear soon, that conditions  necessary to
create such a plasma, 
as depicted in the cartoon (see Fig.~\ref{fig1}), cannot
be fulfilled in the heavy ion collisions. The collision time
is too short, the volume is too small, the temperature is too
low to produce this massless QGP. Therefore the investigations 
developed in the direction, that what is the structure 
of the matter produced in the heavy ion reactions.
(Unfortunately  strongly different structures were also 
named "quark gluon plasma". Thus it is high time to use
different names for the different, well defined matter structures.)  

One of the most important qualifiers to characterize the matter
is the dominant degree of freedom.
In the original quark gluon plasma investigations the dominant 
degrees of freedom were the massless quarks and gluons. 
With the realization, that at the hadronization stage these quarks
interact strongly, the constituent quarks, which have large
effective mass, were considered the dominant degrees of freedom.
This development brought up the idea of coalescence 
hadronization~\cite{alcor}.

\newpage

Here it is important to emphasize, that two different coalescence
 hadronization models were developed.

In our model it was assumed, that during
hadronization both the quark numbers and 
antiquark numbers are conserved~\cite{alcor},
leading to the simple and transparent quark counting scenario
\cite{bialas,alcor2}.
 In the other case one assumes, that 
 new quark - antiquark pairs are created during the hadronization,
and only the net  quark numbers  are conserved
\cite{hwa}.

After these original calculations a large number of new publications,
dealing with different observation, confirmed the validity of the
coalescence model \cite{bialas02,bialas04,greco,fries,molnard}. 
In the present paper we show  that
the most recently found suppression of negative parity baryons
is also the direct consequence of coalescence hadronization.

\section{Introduction }\label{intro}

In the early rehadronization studies the main efforts were concentrated
on the production probabilities of the lowest baryon multiplets. The
structure of the particles belonging to these multiplets were similar:
they all belonged to the spherical symmetric $l=0$  orbital angular momentum
state. Experimentally also these low lying angular momentum states were
observed. Presently, however, an opposite parity state 
($\Lambda(1520)$) also
 has been observed experimentally \cite{na49,star}.

\begin{figure}[tp]
\setlength{\epsfxsize=0.60\textwidth}
\setlength{\epsfysize=0.50\textheight}
\centerline{\epsffile{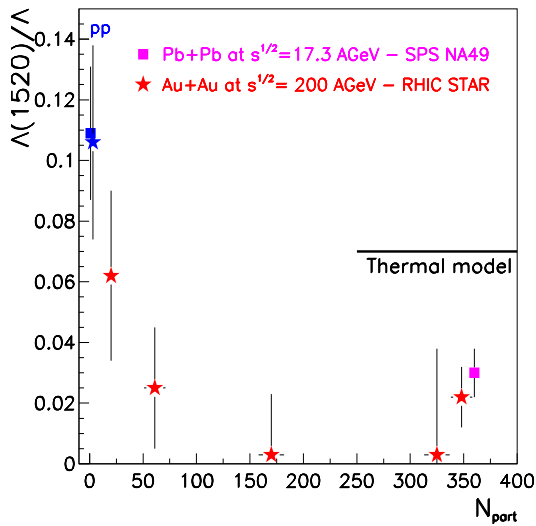}}
\vspace*{-1.5truecm}
\caption[]{ The
$\Lambda (1520)/\Lambda $ ratios measured by
 NA49 (squares) \cite{na49},
and STAR (stars) \cite{star}.
The horizontal line corresponds to a thermal model \cite{braun}.
  }
\label{fig2}
\end{figure}

Figure 2 displays a compilation of experimental values  
 of $\Lambda(1520)/\Lambda$ ratio from NA49~\cite{na49} 
and STAR~\cite{star} together with the result of a thermal model from
Ref.~\cite{braun}.  The figure clearly shows the suppression of this
ratio in central and semi-central $AuAu$ collisions, where
quark coalescence is expected.

In the present paper we  repeat our  earlier calculations of coalescence
of constituent quarks into baryons \cite{alcor}, but now with the inclusion of
opposite parity final states.
We will demonstrate the effect of the symmetry of the internal wave function of the produced hadrons on the transition rates.

In our model the structure of a single hadronization step is assumed
as follows. In the initial state we have a 
diquark (${\cal N}$) interacting with the background 
quark system. Due to this interaction the background quarks form a
screening cluster (${\cal A}$).  
An incident strange quark (${\cal P}$) will pass this cluster 
picking up 
the diquark, forming a new baryon ${\Lambda}$, which leaves 
the reaction zone.

For easier understanding we demonstrate this process  
with  an educational model
and calculate nuclear
cross section for the proton - deuteron pick up reaction:  
\mbox{$ p + (A+n) \rightarrow d + A $} \cite{schiff}. 
 Here the ``proton'' plays the role of the strange quark, 
the ``neutron'' is
the picked up diquark, and the ``deuteron'' is the final state
baryon, $\Lambda$. Real
deuteron has only s-wave and d-wave wave-function component, the
p-wave state is missing. 
Since color forces are much stronger  than
the realistic nuclear forces between real proton and neutron, 
p-wave baryons exist in the nature.
Thus we allow the existence of the p-wave deuteron in
our educational model.

\section{Simple quantum mechanical model for coalescence process}

Considering an incident proton with momentum $ {\bf k_p } $ 
in the center of mass of the $ p $ and $ (A+n) $ system,
the pick-up cross section can be written as follows~\cite{schiff}:
\be
    \sigma_{p + (A+n) \rightarrow d + A}({\bf k_p, k_d}) =
 \frac{v_d}{v_p} | g_{p + (A+n) \rightarrow d + A}({\bf k_p, k_d}) |^2 \ .
\ee
Here the matrix element of the coalescence reaction is determined
by the interaction potential $V_{np}({\bf r_n-r_p})$ and can be
calculated as
\ba
 g_{p + (A+n) \rightarrow d + A}({\bf k_p, k_d}) 
&=& -\frac{M_d}{2 \pi \hbar^2 }  
  \int \displaystyle{\int} \Psi_A^*({\bf\xi})
\phi_{d,l_d}^*({\bf r_n-r_p}) \cdot 
e^{-i{\bf k_d \cdot (r_n+r_p)/2}}\cdot  \NL
&& \hspace*{0.5 truecm}
 \cdot \ V_{np}({\bf r_n-r_p}).  
\Psi_{A,n}({\bf\xi, r_n}) \cdot e^{i {\bf k_p\cdot r_p}} 
d{\bf\xi} d{\bf r_n} d{\bf r_p} \ ,
\ea
The internal wave function of the produced final particle is noted
by $\phi_{d,l_d}$,
where $ l_d $ is the internal angular momentum of the captured neutron 
in the ground and excited state of ``deuteron''.
After integration over variable $\xi$ one obtains
\ba
  g_{p + (A+n) \rightarrow d + A}({\bf k_p, k_d}) &=&    
-\frac{M_d}{2 \pi \hbar^2 }
  \displaystyle{\int}  
\phi_{d,l_d}^*({\bf r_n-r_p})\cdot e^{-i{\bf k_d \cdot (r_n+r_p)/2}} \NL
&& \hspace*{0.5 truecm}  \cdot \ 
 V_{np}({\bf r_n-r_p}) \cdot
\psi_{n}({\bf r_n}) \cdot e^{i {\bf k_p \cdot r_p}}
 d{\bf r_n} d{\bf r_p}   \ .
\ea
Here the wave function of the neutron bound to the nucleus $ A $
is defined as
\be
    \psi_n({\bf r_n}) = \displaystyle{\int} 
\Psi_A^*({\bf\xi}) * \Psi_{A,n}({\bf\xi, r_n}) d{\bf\xi} 
\ee
Introducing new spatial variables 
    $ R = (r_n + r_p)/2 $, \ 
    $ r = (r_n - r_p) $  and the momentum difference
    $ {\bf K} = {\bf k_p - k_d } $,
we  arrive to the expression:
\ba
  g_{p + (A+n) \rightarrow d + A}({\bf k_p, k_d}) &=& 
   -\frac{M_d}{2 \pi \hbar^2 }    
  \displaystyle{\int} \displaystyle{\int}
  F_{d,l_d}^*({\bf k_p, r})   
 \cdot G_{n}({\bf K, R + r/2}) d{\bf r} d{\bf R}  \ \ . \ \ 
\label{integral}
\ea

In the following we calculate this matrix element in eq.(\ref{integral}), 
where the ``deuteron'' and ``neutron'' parts are given as
\ba
 F_{d,l_d}({\bf k_p,r}) &=& \phi_{d,l_d}^*({\bf r})\cdot  V_{np}({\bf r})
 e^{- i {\bf k_p \cdot r/2 }}  \NL
 G_{n}({\bf K, R + r/2}) &=&   \psi_n ({\bf R + r/2})e^{i {\bf K \cdot R }}  
\ea

\subsection{ Deuteron part }
 
We shall assume that the interaction potential between the 
incoming ``proton'' (diquark) and 
the picked up ``neutron'' (quark) has the form:
\be
    V_{np}(r) = \left\{ \begin{array}{cc}  
               V_{inside} & \mbox{for $r < a$} \\
               V_{outside} & \mbox{for $r \geq a$}  \ \ ,
                         \end{array}
                \right.  
\ee   
together with the assumption of $ V_{outside} \longrightarrow \infty  $.
The interaction range $a$ is expected to be in the order of baryon size.
         
\begin{figure}[bp]
\insertplotr{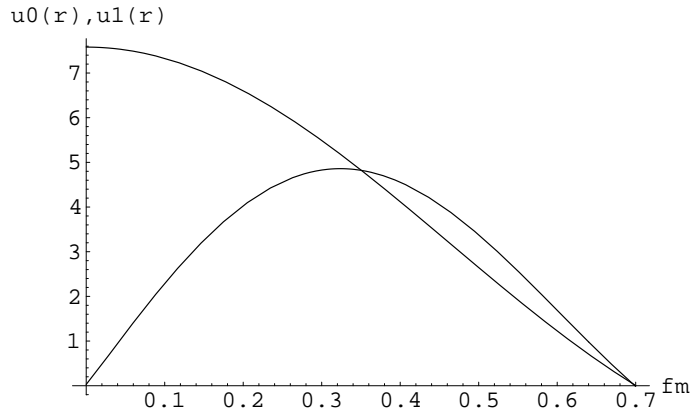}
\vspace*{-0.5cm}
\caption{
Radial wave functions of baryon with $l_d=0$ and $l_d=1$ states  at 
\mbox{$a = 0.7$ fm.} }
\label{figure1}
\end{figure}

 The radial wave function of the deuteron will be  approximated by
 the spherical Bessel functions as
\ba
    u_{d,0}(r) = n_{d,0} \cdot j_0((r/a) \cdot 3.14)   \NL
    u_{d,1}(r) = n_{d,1} \cdot j_1((r/a) \cdot 4.50)  \ , 
\ea
where we used the well known spherical Bessel functions,
\ba
   j_0(z) = sin(z)/z     \NL
   j_1(z) = sin(z)/z^2 - cos(z)/z \ .    
\ea

After the first zero of the Bessel functions,
  $x_0 = 3.14$ and $x_1 = 4.50$,  
we shall assume 
the radial wave function to be identically zero.

Furthermore, 
the normalization equations 
\be
     \int_0^{\infty} u_{d,l_d}(r)^2 r^2 dr = 1
\ee
 can be satisfied by introducing 
$   n_{d,0} = 0.22525 \cdot a^{3/2}$ and
$   n_{d,1} = 0.15327 \cdot a^{3/2}$ 
normalization factors.

The complete deuteron wave function can be written as
\be
   \phi_{d,l_d}({\bf r}) = Y_{l_d,0}(\Theta, \phi) \cdot u_{d,l_d}(r)
\ee

With these notations the "deuteron part" of the matrix element
has the form 
\be
 F_{d,l_d}(k_p,r) = \phi_{d,l_d}({\bf r}) \cdot
 V_{np}(r) \cdot e^{-i{\bf k_p r}/2} \NL
\ee

\subsection{ Neutron  part }
 
The neutron wave function, which is assumed to model quark wave function 
inside the deconfined region, will be approximated by a Gaussian:

\be
   \psi_n({\bf R + r/2 }) = N \cdot \exp \left[-\frac{1}{s^2}({\bf R + r/2 })^2 
   \right]
\label{waven}
\ee
with normalization factor
\be
    N =   \frac{2}{ \pi^{1/4} \cdot s^{3/2} }  \ \ . 
\ee

The Taylor expansion of this wave function around {\bf r = 0} is written as
\ba
\psi({\bf R + r/2 }) &=& \psi({\bf R} )  +
    \nabla \psi({\bf R}) \cdot {\bf (r/2)}   +
      \triangle \psi({\bf R}) \cdot {\bf (r/2)^2}+ ...   
\label{taylorn}
\ea
Substituting the Gaussian wave function from eq.(\ref{waven}) into
eq.(\ref{taylorn}) one obtains
\ba
&& \psi({\bf R + r/2 }) \ \ = \ \   
N \cdot e^{- \frac{1}{s^2}\cdot{\bf R}^2} \cdot \NL 
&& \hspace*{0.5truecm}
\left[ 1 + \frac{-2}{s^2}\cdot{\bf R \cdot (r/2)} + 
\frac{2}{s^4}\cdot(2\cdot(x^2.X^2+y^2.Y^2+z^2.Z^2) - s^2.r^2 )/4 \right] 
\ \ . \ \ \ \ 
\ea

Let us insert this expression into eq.(\ref{integral})
\ba
&& \ \  g_{p + (A+n) \rightarrow d + A}({\bf k_p, K}) = 
  -\frac{M_d}{2 \pi \hbar^2 }    
  \displaystyle{\int} \displaystyle{\int}
  F_{d,l_d}^*({\bf k_p, r}) \ \cdot \  
  N \cdot e^{- {\bf R}^2} \cdot e^{i {\bf K \cdot R }}  
\NL 
&& \hspace*{0.5truecm}  \cdot  
\left( 1 + \frac{-2}{s^2}*{\bf R \cdot r} + 
\frac{2}{s^4}*(2*(x^2.X^2+y^2.Y^2+z^2.Z^2) - s^2.r^2 )\right)
d{\bf r} d{\bf R}  \ \ . \ \ \  
\label{tayint1}
\ea

\subsection{Space integral of the deuteron part }

Let us calculate the following integral:
\ba
I_{l_d}({\bf k_p}) &=&-\frac{M_d}{2\pi \hbar^2} \int F_{d,l_d}^*
({\bf k_p, r}) d{\bf r} \NL
                   &=& -\frac{M_d}{2 \pi \hbar^2 } \int
  Y_{l_d,0}(\theta_r, \phi_r) \cdot u_{l_d}(r) V_{np}(r)      
    \cdot e^{-i{\bf k_p r}/2} d\Omega_r  r^2 dr  
\label{ild}
\ea
Inserting  the Raighley expansion into eq.(\ref{ild})
\ba
 I_{l_d}({\bf k_p}) &=& -\frac{M_d}{2 \pi \hbar^2 } \displaystyle{\int}
  Y_{l_d,0}(\theta_r, \phi_r) \cdot u_{d,l_d}(r) V_{np}(r)    \NL  
    && \cdot 
     \sum_{l=0}^{\infty} \sum_{m=-l}^{l} i^l  j_l(k_p r/2) 
         Y_{l,m}^*(\theta_r,\phi_r)  \cdot Y_{l,m}(\theta_p,\phi_p) 
 d\Omega_r r^2 dr  \ \ , \  \ 
\ea
and using the orthogonality relation 
\be
    \int Y_{l,m}^* (\theta_r,\phi_r) \cdot Y_{k,n} (\theta_r,\phi_r)
   d\Omega_r = \delta_{l,k} . \delta_{m,n}  \ \ , \ \ 
\ee  
we arrive to the following expression:
\ba
 I_{l_d}({\bf k_p}) &=& - i^{l_d} \cdot \frac{M_d}{2 \pi \hbar^2 } 
 \int  u_{d,l_d}(r) V_{np}(r) j_{l_d}(k_p r/2) \cdot r^2 dr 
       \cdot  Y_{l_d,0}(\theta_p,\phi_p) \ \ .
\ea

This integral has to be multiplied by
 the first term of the Taylor expansion
of the "neutron part":
\ba
B_{n}({\bf K})&=&\int N \cdot e^{-(R/s)^2} \cdot Y_{0,0}(\Omega_R)
\cdot e^{i{\bf K.R}} d\Omega_R R^2 dR   \NL
&=&  N\cdot \int e^{-(R/s)^2} \cdot j_0(K \, R) R^2 dR  
\ea

Thus the complete matrix element in first
order approximation is given as:
\be
 g_{p + (A+n) \rightarrow d + A}({\bf k_p, K}) 
= I_{l_d}({\bf k_p})*B_{n}({\bf K})    
\label{tayint2}
\ee

Choosing the Z axis in the direction of $ {\bf k_p}$, 
we have $ \theta = 0, \phi = 0 $, and thus
\ba
   I_{l_d}(k_p) &=& - i^{l_d} [\frac{2l_d+1}{4 \pi}]^{1/2} 
 \cdot \frac{M_d}{2 \pi \hbar^2 }  
 \displaystyle{\int}  u_{d,l_d}(r)
 V_{np}(r)  j_{l_d}(k_p r/2) \cdot r^2 dr  
\label{IL}
\ea

Thus the production rate, $A_{l_d}$ depends on the bombarding momentum, 
$k_p$ and it is determined as
\be
   A_{l_d} (k_p) = {\cal C} \cdot | I_{l_d}(k_p) |^2 \ \ ,
\label{A}
\ee
where ${\cal C}$ is a constant, independent on $l_d$.

\begin{figure}[tp]
\begin{minipage}[t]{6.3cm}
{\insertploth{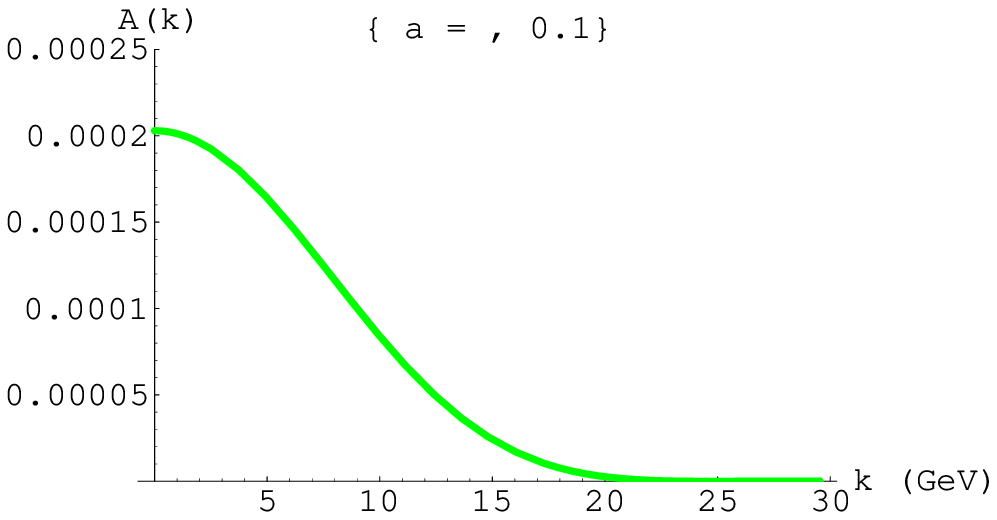}}
\vspace*{-8mm}
\caption[]{
Space integrated matrix \mbox{element} squared $A_{0}(k)$ in eq.(\ref{A})
 at $a=0.3$ fm. }
\label{fig4}
\end{minipage} \hfill
\begin{minipage}[t]{6.3cm}
{\insertploth{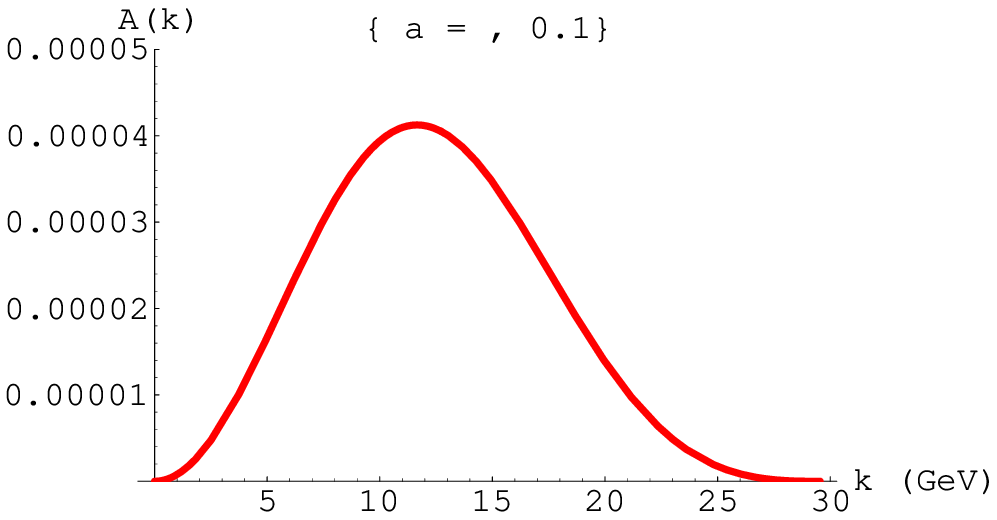}}
\vspace*{-8mm}
\caption[]{
Space integrated matrix \mbox{element} squared $A_{1}(k)$ in eq.(\ref{A})
 at $a=0.3$ fm. }
\label{fig5}
\end{minipage}

\vspace*{2mm}
\begin{minipage}[t]{6.3cm}
{\insertploth{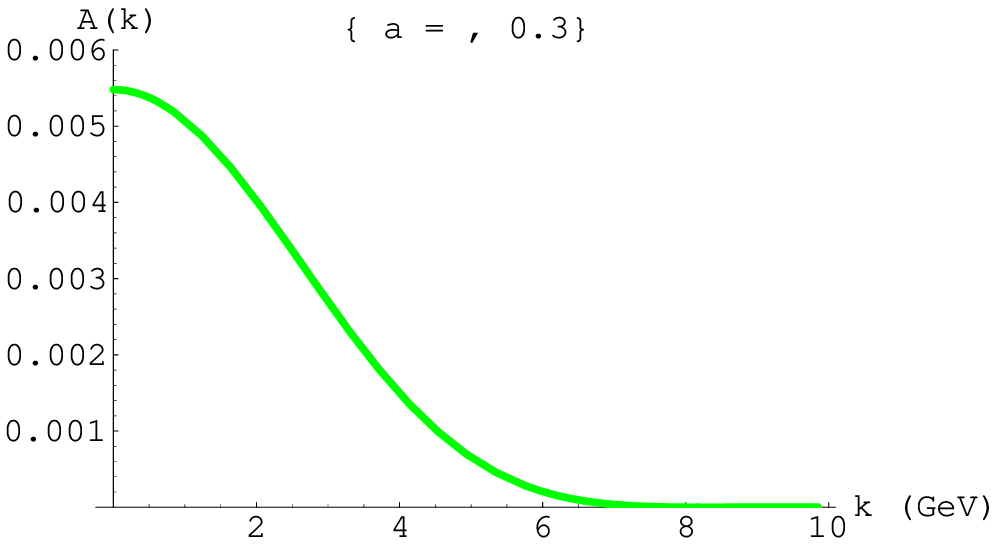}}
\vspace*{-8mm}
\caption[]{
Space integrated matrix \mbox{element} squared $A_{0}(k)$ in eq.(\ref{A})
 at $a=0.3$ fm. }
\label{fig6}
\end{minipage} \hfill
\begin{minipage}[t]{6.3cm}
{\insertploth{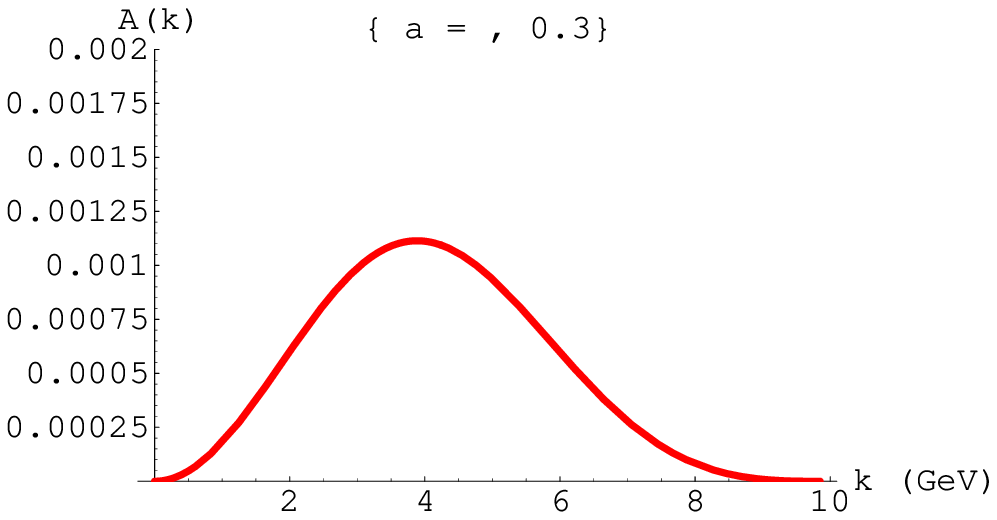}}
\vspace*{-8mm}
\caption[]{
Space integrated matrix \mbox{element} squared $A_{1}(k)$ in eq.(\ref{A})
 at $a=0.3$ fm. }
\label{fig7}
\end{minipage}

\begin{minipage}[t]{6.3cm}
{\insertploth{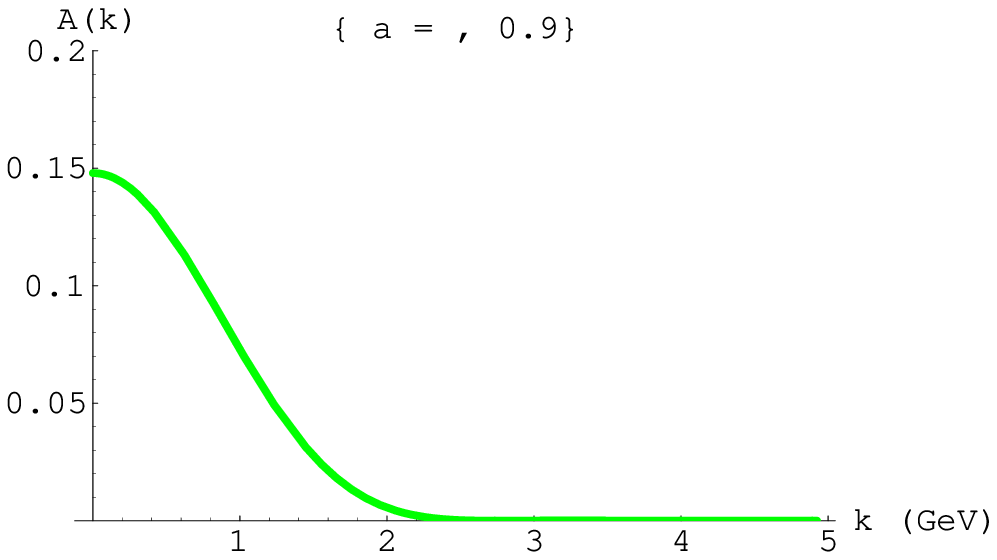}}
\vspace*{-8mm}
\caption[]{
Space integrated matrix \mbox{element} squared $A_{0}(k)$ in eq.(\ref{A})
 at $a=0.9$ fm. }
\label{fig8}
\end{minipage} \hfill
\begin{minipage}[t]{6.3cm}
{\insertploth{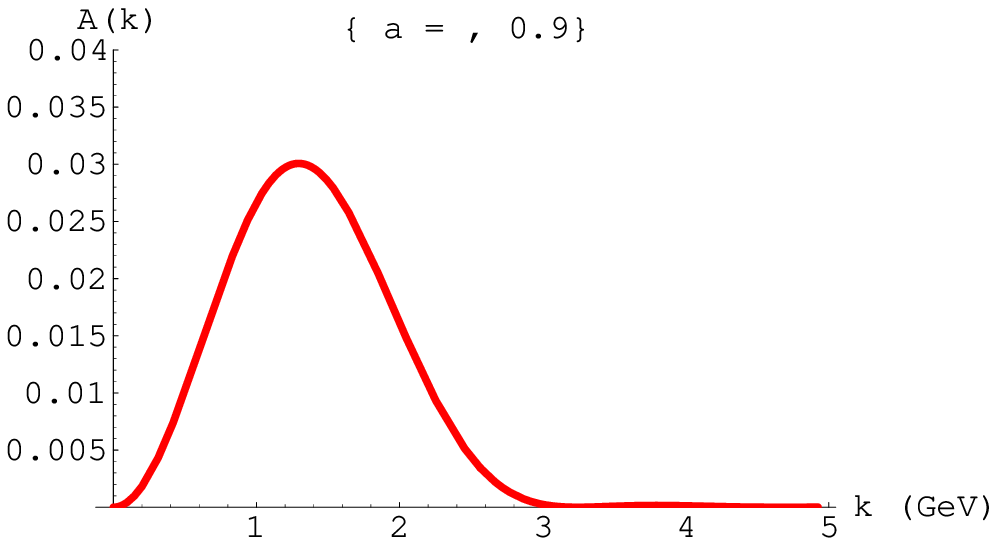}}
\vspace*{-8mm}
\caption[]{
Space integrated matrix \mbox{element} squared $A_{1}(k)$ in eq.(\ref{A})
 at $a=0.9$ fm. }
\label{fig9}
\end{minipage}
\end{figure}

\section{Numerical results}

We calculated the production rates $A_0$ and $A_1$ at interaction 
ranges $a=0.1, \ 0.3, \ 0.9$. The obtained results are displayed in
Figs. 4-9.

The total transition rate, $ R_l$, is obtained by the
integration of $A_{l_d}(k_p)$,
weighted by the distribution function $ f(k_p)$:
\be
     R_{l_d} =  \int_0^{\infty} f(k_p) \cdot A_{l_d}(k_p)  dk_p \ \ .
\label{rat}
\ee  

Here $ f(k_p)$ is the momentum of the proton relative to the
center of mass of the {neutron and screening cluster} system.
For the present numerical calculations we used a Boltzmann
function:
\be
f(k_p) = Exp[-k_p/T] 
\label{boltz}
\ee
(We mention here that for coalescence within parton shower~\cite{hwa04} 
a narrower distribution function should be used.)

\newpage

The obtained $R_0$, $R_1$ production rates can be connected 
to the production rates of $\Lambda(1114)$ and $\Lambda(1520)$,
respectively.
Assuming a fast hadronization at the critical temperature, $T_c$,
one can directly compare the obtained $R_1/R_0$ ratios
to the measured $\Lambda(1520)/\Lambda(1114)$ ratio. Fig. 10
displays our result for $R_1/R_0$ as a function of the interaction 
range, $a$ ({\it solid line}). 
The shaded area indicates the
experimental result $\Lambda(1520)/\Lambda(1114)=0.022 \pm 0.01$
measured by the STAR at RHIC \cite{star}.
The dashed line shows the calculated thermal ratio \cite{braun}.
{}From Fig.10 one can conclude that the $R_1/R_0$ ratio
calculated in our coalescence model using the interaction range 
$a = 0.8-1.0$ fm is consistent with the
experimental data.

\begin{figure}[pt]
{\insertplotg{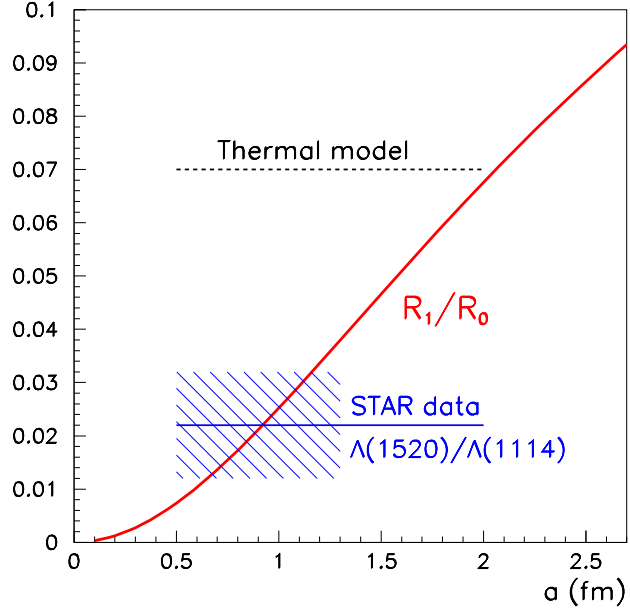}}
\vspace*{-2truecm}
\caption[]{Suppression factor as a function of interaction range, 
$ a$.  }
\label{fig10}
\end{figure}

\section{Summary}

It is an inherent property of the coalescence rehadronization model
that the production of the {$ \Lambda(1520,J^P=\frac{3}{2}^-)$}  baryon
 is strongly suppressed in comparison to the production of 
{ $ \Lambda(1114,J^P=\frac{1}{2}^+)$.} 

This is due to the fact that in  
$\Lambda(1520)$  the orbital angular momentum  of one of the constituent
quark differs by one unit from that of the corresponding quark
in $\Lambda(1114)$.
The strength of the  suppression depends on the length of
 interaction.

{}From the above consideration one may conclude, 
that i) in the STAR Au+Au reaction
a sort of quark matter was formed, meaning that in this case the
dominant degrees of freedom are the constituent quarks,
 and ii) such  matter was not formed in 
the $p + p$ reaction or in peripheral $Au + Au$ reactions. 

One has to mention, however, that it is somewhat
surprising, that
 in the SPS $Pb + Pb$ reaction at $\sqrt{s} = 17.3$ AGeV 
such a suppression also exists.

\section*{Acknowledgment}
The authors acknowledge useful discussions with T.S. Bir\'o 
and L.P. Csernai.

\end{document}